\documentclass[aps,prl,nofootinbib,amsmath,amssymb,amsfonts,twocolumn]{revtex4}

\usepackage{graphicx}
\usepackage{dcolumn}
\usepackage{bm}

\usepackage{amsmath}
\usepackage{latexsym}
\usepackage{epsfig}

\newcommand{\bea}{\begin{eqnarray}}
\newcommand{\eea}{\end{eqnarray}}
\newcommand{\beas}{\begin{eqnarray*}}
\newcommand{\eeas}{\end{eqnarray*}}

\def\bm#1{{\mbox{\boldmath{$#1$}}}}

\begin{document}

\title{Transmission of information via the non-linear Scroedinger equation: \\The random Gaussian input case}
\author{Pavlos Kazakopoulos and Aris L. Moustakas}
\affiliation{Department of Physics, University of Athens, Athens 15784, Greece}
\email{arislm@phys.uoa.gr}
\begin{abstract}
The explosion of demand for ultra-high information transmission rates over the last decade has necessitated the usage of increasingly high light intensities for fiber optical transmissions. As a result, the fiber non-linearities need to be treated non-perturbatively. Similar analyses in the past have focused on the effects of non-linearities on existing transmission technologies, e.g. WDM. In this paper we take advantage of the fact that, under certain assumptions, light transmission through optical fibers can be described using the non-linear Schroedinger equation, which is exactly integrable. As a particular example, we show that in the low Gaussian noise limit, the Gaussian input distribution has a higher mutual information than the transmission using WDM over the same available bandwidth.
\end{abstract}

\maketitle

\section{Introduction}

The possibility of light soliton propagation in a silica fiber was first predicted by Hasegawa and Tappert in 1973 \cite{Hasegawa_Tappert_1973_anomalous,Hasegawa_Tappert_1973_normal}. Since then a tremendous amount of work has been done on both theoretical and experimental aspects of light soliton propagation. Moreover, optical fiber technology has become the centerpiece of
wired telecommunications, with optical fiber networks crossing oceans and webbing continents, carrying the world's digital communications in the form of light soliton pulses.

The property of the silica fiber that makes soliton propagation possible is the Kerr nonlinearity, \emph{i.e.} the dependence of the index of refraction $n$ on the intensity of light $I$:
\begin{eqnarray}
n=n_0(\omega)+n_2 I.
\end{eqnarray}
From this one can straightforwardly derive the equation that governs the propagation of light inside the fiber, the non-linear
Schr\"{o}dinger equation (NLSE):
\begin{eqnarray}
i\frac{\partial u}{\partial t} +  \frac{\partial^2 u}{\partial x^2} + 2\kappa |u|^2 u = 0 \label{nlse}
\end{eqnarray}
where $u(x,t)$ is the envelope of the electric field\footnote{For light propagation inside a fiber, $t$ in (\ref{nlse}) denotes position along the fiber and $x$ denotes time in the comoving frame.} and $\kappa = \pm 1$. The positive value of $\kappa$, describing light propagation in the anomalous dispesion regime, gives the attractive NLSE that admits solitonic solutions
with zero boundary conditions (bright solitons). For $\kappa = -1$ we get the repulsive NLSE, which is valid in the normal dispersion regime and admits solitons for nonvanishing boundary conditions (dark solitons). For either sign of $\kappa$, the NLSE is an integrable hamiltonian system. It belongs to a class of nonlinear equations (together with the KdV equation, the sine-Gordon equations and others) that can be solved exactly by means of the inverse scattering transform (IST) technique (see \emph{e.g.} \cite{Drazin_Johnson_book,Konotop_book} and references therein).

Optical fiber channels are capable of extremely high data transfer rates and constant technological improvements have led to an almost exponential increase in real-world transmission rates. The question naturally arises then, \emph{what is the upper bound imposed on the bit rate  by the physics of the optical fiber, irrespective of any particular technological setup}? The natural framework in which one can address this question is that of information theory, developed by Shannon \cite{Shannon_1948}. In any communication channel, the limiting factor for the rate at which it can carry information is the noise that unavoidably enters along the channel and corrupts the data.  Shannon introduced the concept of \emph{channel capacity}, defined as the maximum possible bit-rate for error-free transmission. The channel capacity $C$ is defined by:
\begin{eqnarray}
C = \mbox{max}_{p_x} \left\{ H[y] - \left\langle H[y|x] \right\rangle_{p_x} \right\} \label{capacity}
\end{eqnarray}
where $y$ is the output (received) signal, $x$ is the input (sent) signal, $p_x$ is the probability distribution of ``symbols" in $x$ (be it letters, fourier components, soliton modes or what it may), and $H$ is the entropy of information:
\begin{eqnarray}
H[x]\equiv - \int dx p_x \log p_x \label{entropy}
\end{eqnarray}
The maximum in (\ref{capacity}) is taken among all possible input distributions $p_x$.
The two quantities that ultimately determine the performance of the channel are the signal to noise ratio (SNR) for the received signal and the bandwidth. For a linear channel (e.g. a copper wire) with additive noise, Shannon's celebrated result \cite{Shannon_1948,Cover_Thomas_book} states that
\begin{eqnarray}
C = W\log \left( 1+\frac{S}{N} \right), \label{shannon}
\end{eqnarray}
$W$ being the bandwidth\footnote{The loss mechanisms for light propagating through silica limit $W$ to a maximum of $50$ THz \cite{Glass_et_al}. Systems in practical use at the moment have a $15$ THz bandwidth.} and $S$, $N$ the average power of the signal and the noise respectively.

However, modern fiber-optics systems
operate in a substantially non-linear regime, rendering the assumption of linearity used to derive (\ref{shannon}) invalid. The additive noise in fiber-optics systems comes from periodically spaced amplifiers (usually erbium-doped segments of fiber) that offset the loss in the electric field amplitude along the fiber\footnote{Besides additive noise, there is also multiplicative noise. This becomes especially important when one considers wavelength division multiplexing (WDM) systems, in which the bandwidth is divided in a multitude of sub-bands (channels). Because of nonlinearity, signals in different channels interact with each other. Because of the statistical independence of signals in different channels one can use a model in which  each channel sees the rest of the bandwidth as multiplicative noise. Very interesting work in the direction of determining the impact of multiplicative noise on the capacity of fiber-optics channels has been done in \cite{Mitra_Stark_2001,Stark_Mitra_Sengupta_2001,Narimanov_Mitra_2002,Green_et_al_2002_XPM,
Turitsyn_et_al_2003_zero_average_dispersion,Kahn_Ho_2004_2004_IT_review}}. These inject noise into the signal, mainly because of amplified spontaneous emission of photons (ASE) \cite{Agrawal_1992,Kaminow_Koch_book}. The initial approach to the effects of noise was to determine how the additive amplifier noise perturbs single solitons, and calculate the jitter introduced into the soliton trains. The result is the so called Gordon-Haus jitter \cite{Gordon_Haus_1986,Kivshar_Haelterman_1994_gordon_haus_dark_solitons}. For both dark and bright solitons, one finds that the perturbation in the frequency (which is also proportional to the velocity of the soliton) is a zero-mean gaussian with variance proportional to the strength of the amplifier noise and the amplitude of the soliton. This  results in a jitter in the soliton arrival times that can cause reading errors at the receiver.  A lot of work has been done on overcoming the restrictions in bit rate due to this effect\footnote{For practical purposes, one needs to obtain a bit error rate lower than $10^{-9}$}. The mechanisms proposed typically involve some form of optical filters or dispersion compensation, such as sliding-frequency filters \cite{Mollenauer_Gordon_Evangelides_1992}, synchronous
modulation \cite{Nakazawa_et_al_1991}, optical phase conjugation \cite{Forysiak_Doran_1995,Goedde_Kath_Kumar_1995,Essiambre_Agrawal_1997_OPC}, and dispersion managed solitons \cite{Karter_et_al_1997,McKinstrie_Santhanam_Agrawal_2002} (there are numerous papers on the subject of suppressing the Gordon-Haus effect, see ch.12 of \cite{Kaminow_Koch_book} for a partial list of references).

This approach is useful when one is considering specific signaling schemes that use soliton trains of fixed amplitude and inter-soliton distance, but to answer the  question of maximum achievable channel capacity we need to abandon the single-soliton assumption and venture out towards a more abstract and general method. In Shannon's theory of linear channels, for an input with a given average power one can show that the the maximal distribution in (\ref{capacity}) is the gaussian. For the nonlinear channel under study, the maximal distribution is much harder to find, but one can still obtain bounds that convey the qualitative behavior of $C$ as a function of the SNR by using a gaussian as input \cite{Mitra_Stark_2001}. Starting with a zero-mean gaussian random electric field with given, constant second moment, we calculate the
density of soliton modes produced by it inside the fiber. The soliton modes are the natural degrees of freedom to use, like the Fourier coefficients would be for a linear channel. The distribution of (either dark or bright) soliton modes is calculated as the density of states (density of eigenvalues) of the linear operator associated with the NLSE in the context of the inverse scattering transform. For the NLSE, the associated linear problem is the $2\times 2$ Zakharov-Shabat eigenvalue problem \cite{Zakharov_Shabat_1972}. The eigenvalue spectrum of the Zakharov-Shabat system includes both continuous and discrete eigenvalues. We are interested in the discrete eigenvalues, because they correspond to soliton modes, and these can be isolated by imposing zero boundary conditions on the eigenstates. One can then use adiabatic perturbation theory based on the IST to compute the statistical uncertainty introduced in the eigevalues by the amplifier noise. This leads to a different distribution at the output, from which the mutual information can be found, and the aforementioned bounds on the capacity $C$ can in principle be computed. The implementation of adiabatic perturbation theory can only be done numerically in the dense soliton limit that we are examining.

\section{Dark Solitons}

\begin{figure}[ht]
\hspace*{-25pt} \vspace*{5pt}
{\includegraphics*[width=0.9\columnwidth]{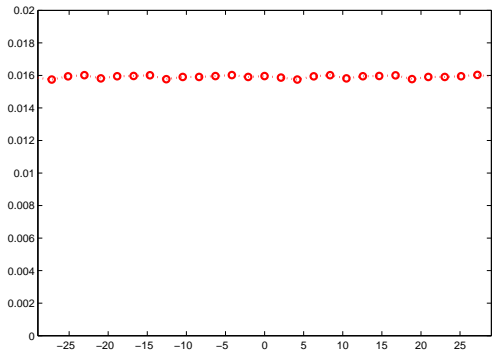}}
\caption{\label{hdosfig}DOS of the hermitian Zakharov-Shabat
eigenvalue problem as a function of the eigenvalue $\lambda$. The
eigenvalues were found by direct diagonalization, using an
adaptation of the modified Ablowitz-Ladik scheme ($D=1$,
$\mbox{size}=20$, $\mbox{step}=0.1$, $200$ runs). We see the
independence of the DOS from $\lambda$.}
\end{figure}

For dark solitons, the associated linear eigenvalue problem is hermitian:
\begin{eqnarray}
\begin{array}{l}
\bm{U_H} \bm{\psi}(x)= \lambda\bm{\psi}(x), \\ \\
\bm{U_H}=\left( \begin{array}{lr} i\frac{\partial}{\partial x} & u^*(x,0) \\ u(x,0) & -i\frac{\partial}{\partial x} \end{array} \right),\\ \\ \bm{\psi}(x)=\left( \begin{array}{c} \psi_1(x) \\ \psi_2(x)  \end{array} \right).
\end{array}
\label{hzs}
\end{eqnarray}
To isolate the soliton modes, we impose the boundary conditions $\psi_1,\psi_2\to 0$ as $x\to \pm \infty$. Hermiticity means that the eigenvalue $\lambda$ is real. For a single soliton, $\lambda$ would give both the velocity and the amplitude of the soliton. In the dense soliton case we are interested in, the eigenvalues are the collective degrees of freedom of the soliton modes. The ``potential" $u(x,0)$ is the initial condition of
the NLSE (\ref{nlse}) or, in the case of optical fibers, where `$x$' denotes time, the envelope of the electric field that goes into the fiber.  We choose $u$ to be gaussian random, i.e.
\begin{eqnarray}
u(x)= \frac{1}{\sqrt{2}}\left(u_1(x)+i u_2(x)\right) \label{gauss}
\end{eqnarray}
with $u_1,u_2$ real, $\left\langle u_i(x) \right\rangle = 0$,
and $\left\langle u_i(x)u_j(x')\right\rangle=D\delta_{ij}\delta(x-x')\;$,$\;\;i,j=1,2$. The constraint on the second moments of $u_i$
translates to a power constraint for the ingoing electric field, specifically that it has average power $D$. The DOS of this operator with these boundary conditions has been known in the literature for some time \cite{Ovchinnikov_Erikhman_1977,Hayn_John_1987,Gredeskul1990_Dark_Solitons,Bartosch_thesis}. The DOS is constant, independent of $\lambda$. In FIG.~\ref{hdosfig} we see the results of numeric simulations for the DOS. The simulation was done using an adaptation of the modified Ablowitz-Ladik scheme \cite{Weideman_Herbst_Modified_Ablowitz_Ladik} for the hermitian problem.
Beyond determining the DOS, we want to know whether the eigenvalues are statistically independent,
\emph{i.e.} whether $\rho(\lambda,\lambda ') = \rho(\lambda)\rho(\lambda ')$. We were able to show this both analytically and numerically. On the analytical side, we used Halperin's method \cite{Halperin_1965_DOS,Frisch_Lloyd_1960}. We define the variables $\theta(x)=\mbox{arg}\left( \frac{\psi_1(x)}{\psi_2(x)} \right)$, $\theta '(x)=\mbox{arg}\left( \frac{\psi_1 '(x)}{\psi_2 '(x)} \right)$. Their evolution along $x$ is a Markov process and from (\ref{hzs}) we can derive the Fokker-Planck equation for their probability distribution $P(x;\theta,\theta ')$:
\begin{eqnarray}
\frac{\partial P}{\partial x} &=& \lambda \frac{\partial P}{\partial \theta} + \lambda' \frac{\partial P}{\partial \theta'} +
D\left(\frac{\partial^2 P}{\partial \theta^2} + \frac{\partial^2 P}{\partial {\theta}^2} \right) \\ \nonumber
&+& 2 D \frac{\partial^2}{\partial \theta \partial \theta'}\left( \cos(\theta - \theta')P\right). \label{fokkerplanck}
\end{eqnarray}
We also derive the equation for the quantity
$F(x;\theta \theta') \equiv \left\langle \frac{\partial \theta}{\partial \lambda} \frac{\partial \theta '}{\partial \lambda '} \delta(\theta (x)-\theta) \delta (\theta ' (x)-\theta ') \right\rangle$ (the brackets denote averaging over the gaussian ensemble of $u$'s)
\begin{eqnarray}
 \frac{\partial F}{\partial x} &=& -H -G + \lambda \frac{\partial F}{\partial \theta} + \lambda' \frac{\partial F}{\partial \theta'} \\ \nonumber
 &+& D\left(\frac{\partial^2 F}{\partial \theta^2} + \frac{\partial^2 F}{\partial {\theta}^2} \right)
 + 2 D \cos(\theta - \theta') \frac{\partial^2 F}{\partial \theta \partial \theta'} \label{sourcefokkerplanck}
\end{eqnarray}
where $H(x;\theta,\theta ')$, $G(x;\theta,\theta ')$ satisfy the same equation as $P(x;\theta,\theta ')$ but with $-P$ as an extra source term. $F$ grows as $\rho(\lambda,\lambda ')\times x^2$ for large $x$ and from (\ref{fokkerplanck}),(\ref{sourcefokkerplanck}) we are able to calculate
$\rho(\lambda,\lambda ')$ and show that it factorizes.

\begin{figure}[ht]
\hspace*{-25pt} \vspace*{5pt}
{\includegraphics*[width=0.9\columnwidth]{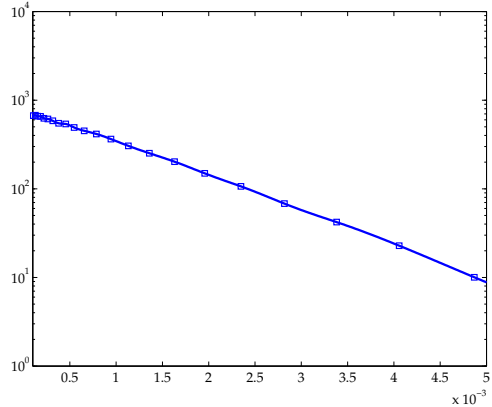}}
\caption{\label{distfig} Logarithmic plot of the distribution of
the distance between successive eigenvalues of the hermitian
Zakharov-Shabat operator ($D=1$, $\mbox{size}=100$,
$\mbox{step}=0.1$, $100$ runs). The linearity of the data shows
that the distances follow a Poisson distribution, i.e. there is no
bias in the magnitude of the distance, in agreement with our
theoretical result about statistical independence of eigenvalues
even in the dense soliton limit.}
\end{figure}

Numerically, we find the distribution of the distances between neighboring eigenvalues. Statistical independence means that the distances must follow a Poisson distribution, and this is indeed what we find. The results are shown in FIG.~\ref{distfig}

The effect of (weak) additive noise coming from periodic amplification on the eigenvalues can be studied in the context of adiabatic perturbation theory and the IST \cite{Lashkin_2004_dark_soliton_perturbation}. It is thus shown that for white gaussian noise, the disturbance of the eigenvalues is gaussian, with zero mean and a variance $\left \langle \Delta \lambda^2 \right \rangle$ proportional to the strength of the noise and the \emph{inverse participation ratio} $\int dx ||\psi||^4$ of the corresponding (normalized) eigenstates. The inverse participation ratio (IPR)in the many soliton situation that we are interested in is beyond analytical treatment. In the next two figures we show the results of numerical simulations where we calculated the IPR for the eigenstates of $\bm{U_H}$ found using direct diagonalization of $\bm{U_H}$ with a central differences discretization (in order to impose zero boundary conditions at the edges).

\begin{figure}[ht]
\hspace*{-25pt} \vspace*{5pt}
{\includegraphics*[width=0.9\columnwidth]{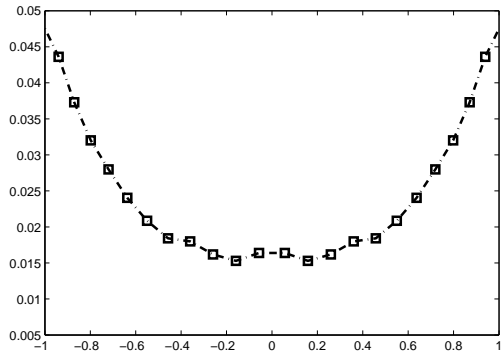}}
\caption{\label{psi_4}Inverse participation ratio as a function of
the eigenvalue $\lambda$, for a given average power of the input
($D=1$, $\mbox{size}=100$, $\mbox{step}=0.1$, $100$ runs).}
\end{figure}

\begin{figure}[ht]
\hspace*{-25pt} \vspace*{5pt}
{\includegraphics*[width=0.9\columnwidth]{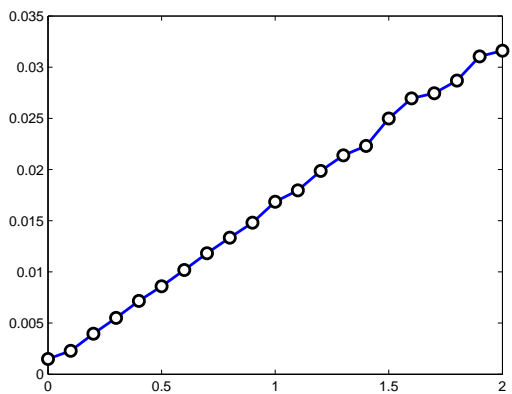}}
\caption{\label{loclength}Inverse participation ratio, averaged
over the eigenvalues, as a function of the average input power $D$
($\mbox{size}=100$, $\mbox{step}=0.1$, $100$ runs).}
\end{figure}

In FIG.~\ref{psi_4} we see how the IPR (locally averaged for smoothness) behaves as a function of the eigenvalue $\lambda$ for a given $D$. We are actually interested only in the middle, flat section, because the raising of the edges is an artefact of the central difference method that creates an over-concentration of eigenvalues near the edges of the spectrum.
Averaging over this flat segment for different $D$'s, we see in  FIG.~\ref{loclength} the dependence of $\int dx ||\psi||^4$ on the average input power $D$. The linearity indicates that the IPR is actually proportional to the localization length (the inverse Lyapunov exponent) of $U_H$ \cite{Bartosch_thesis}.

\section{Bright Solitons}

The non-hermitian Zakharov-Shabat eigenvalue problem is defined by the system of equations
\begin{eqnarray}
\begin{array}{l}
\bm{U} \bm{\psi}(x)= z\bm{\psi}(x), \\ \\
\bm{U}=\left( \begin{array}{lr} i\frac{\partial}{\partial x} & u^*(x,0) \\ -u(x,0) & -i\frac{\partial}{\partial x} \end{array} \right),\\ \\ \bm{\psi}(x)=\left( \begin{array}{c} \psi_1(x) \\ \psi_2(x)  \end{array} \right).
\end{array}
\label{nhzs}
\end{eqnarray}
on the infinite line, together with the boundary conditions $\psi_1,\psi_2\to 0$ as $x\to \pm \infty$. The star denotes complex conjugation, and the eigenvalue $z=\xi+i\eta$ is generally complex. As in the hermitian case, the ``potential" $u(x,0)$ is the envelope of the electric field that goes into the fiber. We want to find the density of states (DOS) of this operator when $u$ is gaussian random, as in (\ref{gauss}).  The DOS $\rho(\xi,\eta)$ of this random operator determines the entropy of information carried by the gaussian signal through the formula $I=-\int d\xi d\eta\rho(\xi,\eta)\log\rho(\xi,\eta)$.

Contrary to its hermitian counterpart, this DOS is not known in the literature. Halperin's method \cite{Halperin_1965_DOS}, which works so nice for the hermitian/dark soliton case, fails here because of non-hermiticity. Non-hermitian
random operators have received considerable attention in the literature (see \cite{Marchetti_Simons_1999_tail_states} for a list of important references on the subject). They have a wide range of applications, in non-equilibrium statistical mechanics, random classical dynamics, the physics of polymers, QCD, neural networks, and, in the case at hand, soliton physics and communications. Non-hermiticity means that the eigenvalues migrate to the complex plane. This complicates the computation of their statistical properties - most notably their DOS - relative to hermitian operators. The underlying reason for this is that the propagator in the hermitian case is analytic except on branch cuts of the real axis, where the real eigenvalues condense. This introduces constraints that facilitate the computation of the DOS. When the eigenvalues are complex this is no longer true. In many cases however one can obtain approximate results for the DOS. The hermitization method, developed by Feinberg and Zee in \cite{Feinberg_Zee_Hermitization} can be applied here. The method gives the self-consistent Born approximation (SCBA) to the density of states. Instead of looking  at $\bm{U}$ directly, one starts with a ``hermitized" operator
\begin{eqnarray}
\bm{H} = \left( \begin{array}{lr} 0 & \bm{U}-z \\ & \\ \bm{U}^{\dagger}-z^* & 0 \end{array} \right)
\end{eqnarray}
and calculates the propagator $\bm{G}=\frac{1}{\mu-\bm{H}}$ in the SCBA. The DOS is then given by the derivative with respect to $z^*$ of the trace of the lower-left block of $\bm{G}$. We omit the details of the calculation and state only the final result.
We find the DOS to be uniform inside a uniform band centered around the $\xi$ axis on the complex $z$ plane, with the width being proportional to $D$:
\begin{eqnarray}
\rho(\xi,\eta) = \left \{ \begin{array}{cl} \frac{1}{2\pi D}, & |\eta|\leq D \\ & \\ 0, & |\eta|>D \end{array} \right.
\end{eqnarray}

To move beyond the SCBA, and see how the DOS ``frays" near the edges of the band, we used the method of optimal fluctuations
\cite{Zittartz_Langer_1966,Izyumov_Simons_1999_optimal_fluctuations}.   The basic idea is that states outside the band given by the SCBA are created by atypically strong fluctuations
of the ``potential" $u(x,0)$. Such fluctuations occur with an exponentially small probability $\exp\left[ -W/D \right]$, where
\begin{eqnarray}
W = \frac{1}{2}\int dx |u(x,0)|^2. \label{functional}
\end{eqnarray}
Minimizing the functional $W[u]$ in (\ref{functional}) with respect to $u$, with the additional constraint $\det (\bm{U}-z)=0$ enforced as a Lagrange multiplier, we can determine the density of states outside the band with exponential accuracy. The minimization involves the solution of a pair of coupled nonlinear ordinary differential equations and the result for the optimal potential is:
\begin{eqnarray}
u(x) = 4i\eta\left( \frac{e^{2\eta x}+e^{-2\eta x}}{e^{4\eta x}+e^{-4\eta x}+2} \right)e^{-2i\xi x}. \label{optimal}
\end{eqnarray}
From (\ref{functional},\ref{optimal}) we then find that to exponential accuracy, the DOS for $|\eta| \gg D$ is
\begin{eqnarray}
\rho(\xi,\eta) \sim e^{-\frac{4|\eta|}{D}}.
\end{eqnarray}

\begin{figure}[ht]
\hspace*{-25pt} \vspace*{5pt}
{\includegraphics*[width=0.9\columnwidth]{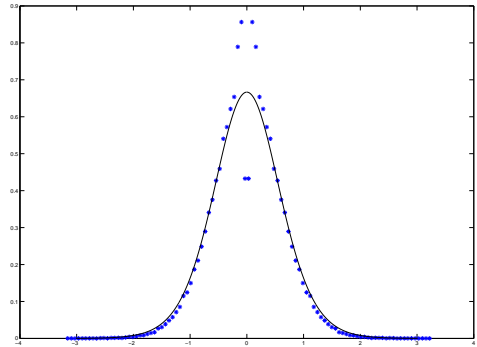}}
\caption{\label{dosfig}Theoretical curve (solid line) and results
of numerical simulations for the profile of the DOS vs. $\eta$. We
have used the modified Ablowitz-Ladik diagonalization scheme
\cite{Weideman_Herbst_Modified_Ablowitz_Ladik}. The value of $D$
is $1$, the size of the system is $L=135$ and the step size is
$0.075$. The disturbance near $\eta=0$ is a finite-size effect.
The localization length $l=\lambda^{-1}$ grows as $l\sim
3D/2\eta^2$ near $\eta=0$ and so numerical results become
unreliable for $|\eta| \lesssim \sqrt{D/L}$ ($\sim 0.1$ here).}
\end{figure}

Although these approximate methods provide some measure of knowledge of the DOS (and consequently the entropy of information of the gaussian random signal), an exact expression is always desirable, for obvious reasons. We were able to obtain such a result using
a variation of the Thouless formula \cite{Thouless_1972}, that relates the DOS with the Lyapunov exponent of the operator  $U$
in (\ref{nhzs}). The Lyapunov exponent $\lambda$ measures the rate of exponential increase of the modulus of a typical solution of (\ref{nhzs}) as one goes towards larger $x$ with no boundary conditions are imposed. One possible definition is:
\begin{eqnarray}
\lambda = \lim_{x\to\infty}\frac{1}{x}\ln\left(|\psi_1(x)|^2+|\psi_2(x)|^2\right)^{\frac{1}{2}}.
\end{eqnarray}
We were able to show that for any set of finite boundary conditions, a unique, positive Lyapunov exponent exists and it is related to the DOS by the formula
 \begin{eqnarray}
\rho(\xi,\eta) = \frac{1}{2\pi}\left( \frac{\partial^2}{\partial\xi^2}+\frac{\partial^2}{\partial\eta^2}\right)\lambda(\xi,\eta).
\label{thoulessformula}
\end{eqnarray}
The eigenvalue problem (\ref{nhzs}) has enough symmetry to make
the exact calculation of $\lambda$ (and from this, $\rho$)
possible. The details of this calculation are contained in the
attached paper (together with the detailed proof of
(\ref{thoulessformula}) and the relevant references) which has
been published in \emph{Phys. Rev. E}
\cite{Kazakopoulos2008_NLSE}. The expression we arrive at for the
DOS is:
\begin{eqnarray}
\rho(\xi,\eta) = \frac{2}{\pi D} \frac{\frac{2\eta}{D}\coth\left( \frac{2\eta}{D} \right) -1}{\sinh^2\left(\frac{2\eta}{D}\right)}
\label{nhdos}
\end{eqnarray}
We have also been able to show that (\ref{nhdos})remains valid for large $\xi$ in the case of a purely real gaussian random potential that maximally breaks rotational symmetry (completely polarized input).
The comparison of this prediction with numerical simulations is shown in FIG.~\ref{dosfig}.

This description of the input signal in terms of the density of $\xi$ and $\eta$
is complete only in the cases where one can neglect
the effects of jitter in the positions of the solitons and concentrate on the
shift in the eigenvalues as the primary cause of signal distortion. If we want to include
the positional jitter, we must take into account the information contained in the
positions of the solitons. This is done by considering another set of complex numbers
$b_z$, one for each solitonic excitation, which are associated with soliton positions
\footnote{In the simplest case of a single localized eigenstate with $z=\xi
+ i\eta$ and $b_z=|b|e^{i\phi}$, the corresponding soliton has
amplitude $\xi$, velocity $2\eta$, initial ``position''
$t_0=\frac{\ln |b|}{2\eta}$ and initial phase $\phi_0=\phi-\pi$
\cite{Konotop_book}.}
and together with the eigenvalues fully determine  the solitonic part
of the signal. Since we study the effects of noise on a random Gaussian
input pulse, where all eigenstates are localized in the infinite pulse
duration limit, the eigenvalues $z$ and and the corresponding $b_z$
give a complete description of the signal.

We were able to compute the distribution of $b_z$ resulting from a random
gaussian input analytically by using the definition of $b_z$ and expressing
it in terms of the behavior of the eigenfunctions at the edges of the pulse,
which in turn we can express as sums of random variables with calculable distributions.
More specifically, the precise asymptotic conditions for the eigenfunctions are:
\begin{eqnarray} %
 \begin{array}{l} \bm{\Psi}_z(x) \rightarrow \left ( \begin{array}{c} 0 \\ 1 \end{array} \right )
  e^{izx},\;\; \bar{\bm{\Psi}}_z(x) \rightarrow \left ( \begin{array}{c} 1 \\ 0 \end{array} \right )
  e^{-izx} \;\;\;\; \mbox{as} \; \; x \rightarrow \infty \\ \\
  \bm{\Phi}_z(x) \rightarrow \left ( \begin{array}{c} 1 \\ 0 \end{array} \right )
  e^{-izx}, \; \bar{\bm{\Phi}}_z(x) \rightarrow \left ( \begin{array}{c} 0 \\ 1 \end{array} \right )
  e^{izx} \;\;\;\; \mbox{as} \; \;  x \rightarrow -\infty. \end{array}%
\label{zs_asymptotic}
\end{eqnarray} %
where the two sets of solutions are related through an S-matrix:
\begin{eqnarray}
\left[\begin{array}{c} %
\bm{\Phi}_z(x) \\
\bar{\bm{\Phi}}_z(x) \end{array} \right] =
\left(\begin{array}{cc} %
b(z) & a(z) \\ \bar{a}(z) & \bar{b}(z)
\end{array}
\right)
\left[\begin{array}{c} %
\bm{\Psi}_z(x) \\
\bar{\bm{\Psi}}_z(x) \end{array} \right]
\label{linear_1}
\end{eqnarray}
with the $a$, $b$'s being the transmission and reflection
coefficients respectively.  By taking into account the
symmetry of the problem under complex conjugation
it is possible to show that $a(z^*)
=\bar{a}^*(z)$ and $b(z^*) = -\bar{b}^*(z)$, where the star
($^*$) denotes the complex conjugate.

When the above solutions correspond to a localized eigenfunction
with eigenvalue $z$, the transmission coefficient $a(z)$ has to
vanish at that $z$,
making the two sets of solutions directly proportional:
\begin{eqnarray}
\begin{array}{c}
\bm{\Phi}_{z}(x) = b_z\bm{\Psi}_{z}(x) \\ \\
\bar{\bm{\Phi}}_{z}(x) = -b_z^*\bar{\bm{\Psi}}_{z}(x)
\end{array}
\label{linear_2}
\end{eqnarray}
where $\Phi_z$ and $\bar{\Phi}_z$ are the admissible
exponentially decaying eigenfunctions for $Im(z)>0$ and $Im(z)<0$,
respectively. Then, for $\eta > 0$ we have from
(\ref{zs_asymptotic},\ref{linear_2})
\begin{eqnarray}
\begin{array}{c}
\bm{\Psi}_z(x) \rightarrow \left ( \begin{array}{c} 0 \\ 1 \end{array} \right )
  e^{izx} \; \mbox{as} \; x \rightarrow \infty \\ \\
\bm{\Psi}_z(-x) \rightarrow \left ( \begin{array}{c} b_z^{-1} \\ 0 \end{array} \right )
  e^{izx} \; \mbox{as} \; x \rightarrow \infty
\end{array}
\end{eqnarray}
Defining $\tilde{\bm{\Psi}}_z(x)\equiv \bm{\Psi}_z(-x)$ we can write:
\begin{eqnarray}
b=\lim_{x\to\infty}b(x),\;\;\; b(x)\equiv\frac{\psi_2(x)}{\tilde{\psi}_1(x)},
\end{eqnarray}
where for convenience we have dropped the subscript $z$.
The time evolution of $\ln{b}$ is found from (\ref{nhzs}),
\begin{eqnarray}
\frac{\partial\ln{b}}{\partial x} = i \left ( uf + \tilde{u}^* \tilde{f} \right )
\label{lnb_evolution}
\end{eqnarray}
with
$f(x) = \frac{\psi_1}{\psi_2}$, $\tilde{f}(x) =
\frac{\tilde{\psi}_2}{\tilde{\psi}_1}$, $\tilde{u}(x) \equiv
u(-x)$, and
\begin{eqnarray}
\begin{array}{l}
\frac{\partial f}{\partial x} = -2izf + i u^* - iuf^2  \\ \\
\frac{\partial \tilde{f}}{\partial x} = -2iz\tilde{f} - i \tilde{u} + i\tilde{u}^*\tilde{f}^2.
\end{array}
\label{fg_evolution}
\end{eqnarray}

Eqs. (\ref{lnb_evolution},\ref{fg_evolution}) can be used to write
down a Fokker-Planck equation for the probability distribution
of $\ln b_z$. The stationary solution of this equation is the actual
distribution of $\ln b_z$ in the infinite pulse limit, and it turns
out to be uniform, as we would expect from the translational invariance
of the input signal. The precise way in which this happens can be
found by using (\ref{lnb_evolution}) to express $\ln b_z$ as a sum
of random variables with known limit distribution. The details of this calculation
together with the relevant references are included in the attached paper.
The result is a Gaussian distribution for the real part of $\ln b_z$
in the case of an unpolarized incoming pulse, with zero mean and variance growing as:
\begin{eqnarray}
\sigma^2 = 4\sqrt{\pi}\frac{\eta e^{\frac{2\eta}{D}}}
{\sinh\left ( \frac{2\eta}{D}\right )}T\ln\frac{T}{2\tau},
\label{lnb_variance}
\end{eqnarray}
where $T$ is the duration of the incoming pulse and $\tau$
is the inverse bandwidth.
The imaginary part of $\ln b$ is an angle and so, although it
follows the same distribution as the real part, will due to
periodicity become uniformly distributed in $[0,2\pi)$.

For a completely polarized incoming pulse,
the distribution of $\ln|b_z|$ will asymptotically follow a
Cauchy distribution scaling like $T/\tau$. Its statistical median
will be zero by symmetry. The phase of $b$ will again be uniform over
$[0,2\pi)$, although the mechanism is different in this case.
For the special case of $\xi=0$, the scale
parameter of the Cauchy distribution can be calculated more explicitly to be
\begin{equation}
\gamma \sim \frac{e^{\eta /D}}{I_0(\eta/D)}\frac{T}{\tau}
\end{equation}
where $I_0$ is a modified Bessel function of the first kind.

\section{Optical Fiber Channel Capacity}

Armed with the knowledge of the distributions of scattering data
$\{z,b_z\}$ generated by a random Gaussian input signal, we
address the question of the capacity, or spectral efficiency, of
the optical fiber channel. Historically, there has been a steady
exponential increase in the transmission capacity of fiber optics
communications systems, resulting mainly from higher transmission
rates and the implementation of wavelength division multiplexing
(WDM). WDM systems generally use binary on-off keying as their
signaling scheme, thus limiting the spectral efficiency (bits
transmitted per second per Hz of bandwidth) to 1 b/s/Hz. More
complicated signaling schemes can go beyond this limit, and will
eventually be required in order to make use of the potential of
the optical fiber network. Some amount of interesting work has
been done on the question of the capacity limits for WDM and
related systems, both analytical and numerical (see
\cite{Mitra_Stark_2001,
Stark_Mitra_Sengupta_2001,Narimanov_Mitra_2002,Green_et_al_2002_XPM,
Turitsyn_et_al_2003_zero_average_dispersion,Kahn_Ho_2004_2004_IT_review,
Djordjevic_2005,Ivcovic_2007,Mecozzi_2001,Mecozzi_1994,Peddanarappagari_1997,Ho_2005}
and references therein). In particular, in \cite{Mitra_Stark_2001}
the spectral efficiency predicted is about $8.5bits/sec/Hz$, in
the absence of jitter. In our model we consider a single-channel
mode instead of the multi-channel setup of WDM. This has the
advantage of avoiding the multiplicative noise resulting from
interference between different channels. The modulation of the
very large bandwidth of optical fibers ($\sim$ 50 THz) is beyond
the ability of current electronic equipment, but we are interested
in the limits posed by the very physics of nonlinearity, which
future technological advances should be able to approach.

The use of a definite input signal (random Gaussian) provides a \emph{lower
bound} for the capacity of the system, since the Gaussian is not
necessarily the optimal input distribution for nonlinear channels.
As we said in the introduction, the method we use is a transformation
to the field of eigenvalues of the Zakharov-Shabat operator, where
the amplifier noise becomes additive. We consider a system
consisting of a fiber span of total length $L$, and $N_a$ equally
spaced identical amplifiers that compensate for the loss of strength in the signal
while unavoidably injecting it with white noise (amplified spontaneous emission noise
or ASE noise). For long-range transmission systems $L\sim 1000 \mbox{km}$
and the inter-amplifier spacing is $L_a\sim 50-100 \mbox{km}$. We study
the case of bright solitons only, since dark solitons are not used
for large distance communications. Mathematically
it is modeled by the addition of a term $i\sum_{n=1}^{N_a}\delta\left(x-nL_a\right)f_n(t)$
on the rhs of Eq. (\ref{nlse}) with the roles of space and time interchanged.
The delta-correlated white noise added at each amplifier has noise strength \cite{Moore2006_LargeNLSE_Perturbations}
\begin{equation}
\left< f_n(t)f_m(t')\right> = \sigma^2\delta_{nm}\delta(t-t'),\;\;\;
\sigma^2 = \frac{h\nu_0\eta_{sp}(G-1)^2}{G\ln G},
\label{add_noise}
\end{equation}
where $h$ is Planck's constant, $\nu_0$ is the carrier frequency, $\eta_{sp}$
is the spontaneous emission factor and $G = e^{\alpha L_a}$ is the amplification
factor ($\alpha=0.2\mbox{dB/km}$). The noise causes a random shift in the
eigenvalues:
\begin{equation}
z_{\mbox{out}} = z_{\mbox{in}} + \delta z.
\label{eig_shift}
\end{equation}

As long as the signal to noise ratio, which can be increased by increasing the input power
or reducing the distance between amplifiers, is large, $\delta z$ (or rather its variance)
can be calculated using adiabatic perturbation theory:
\begin{equation}
\delta z = \frac{1}{2i}\frac{\int \left(f\psi_1^2 + f^*\psi_2^2\right)}{\int \psi_1\psi_2}.
\end{equation}
The denominator here is the non-Hermitian norm of the eigenvector. The
second moments of $\delta z$, $\left<\delta z^2\right>$ and $\left<|\delta z|^2\right>$
contain integrals of the form $\int \psi_1^2\psi_2^2$ that do not correspond to integrals
of the motion and it is hard, if at all possible, to calculate them analytically.
They can however be evaluated numerically, using the eigenstates corresponding
to each eigenvalue \footnote{Note that in our setup this is true only for a random
\emph{Gaussian} incoming pulse, because $\int|u^2|$ is an integral of the motion
for the NLS equation, and so the statistics are not altered by propagation. A different
initial distribution of signals would require Optical Phase Conjugators to periodically
restore the statistics.}.

The rate (per Hz) of information transmission in our model (ignoring jitter for the moment) is given by:
\begin{eqnarray}
R =  H[z_{out}] - \left\langle H[z_{out}|z_{in}] \right\rangle_{p_{z_{in}}}.
\label{rate}
\end{eqnarray}
Since we work in the perturbative regime, we can consistently approximate
$H[z_{out}]$ with $H[z_{in}]$, and because the noise is additive,
$H[z_{out}|z_{in}] = H[\delta z]$. Thus we arrive at a simplified formula
for the rate:
\begin{equation}
R = H[z_{in}] - H[\delta z].
\end{equation}
The entropy of $z_{in}$ is calculated from the distribution of eigenvalues
given in the previous section: $P(\xi,\eta) = P_{\xi}(\xi)P_{\eta}(\eta)$.
Only positive values of $\xi$ and $\eta$ correspond to independent eigenstates,
the rest being ``mirror images'' of these \footnote{The eigenstates for $\eta<0$
are related to those with $\eta>0$ simply by complex conjugation of the
ZS eigensystem, while those with $\xi<0$ have positive counterparts with $\xi'=\xi + \frac{\pi}{\tau}$,$\tau$
being the inverse bandwidth.}. $\xi$ is uniformly
distributed, with maximum value $\Lambda = 2\pi B$, where $B$ is the bandwidth,
and $\eta$ is distributed according to:
\begin{eqnarray}
P_{\eta}(\eta) = \frac{4}{D} \frac{\frac{2\eta}{D}\coth\left( \frac{2\eta}{D} \right) -1}{\sinh^2\left(\frac{2\eta}{D}\right)}.
\label{etados}
\end{eqnarray}
The incoming entropy (per symbol,in \emph{nats}) then is:
\begin{equation}
H[\xi,\eta] = H[\xi]+H[\eta] = \ln \Lambda + \ln\left(\frac{D}{4}\right) + 0.08\;.
\label{incoming_entropy}
\end{equation}

\begin{figure}[ht]
\hspace*{-25pt} \vspace*{5pt}
{\includegraphics*[width=0.9\columnwidth]{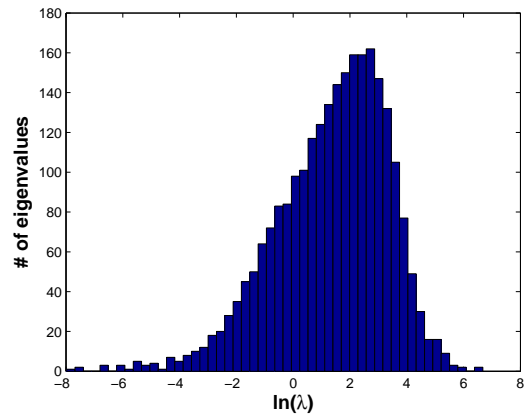}}
\caption{\label{cov_eig_distr}Distribution of the logarithm of the
eigenvalues of the covariant matrix. The value of $D$ is $2$, the
size of the system is $L=150$ and the step size is $0.06$. }
\end{figure}

The entropy of the noise is computed in the Gaussian approximation
\footnote{Note that this makes our lower bound computation more solid,
since the Gaussian has maximum entropy for given variance.},
as the logarithm of the covariance matrix of $\delta\xi,\delta\eta$.
The distribution of the logarithm of eigenvalues of this matrix for a value of
$D = 2$ is shown in FIG.~\ref{cov_eig_distr}.

\begin{figure}[ht]
\hspace*{-25pt} \vspace*{5pt}
{\includegraphics*[width=0.9\columnwidth]{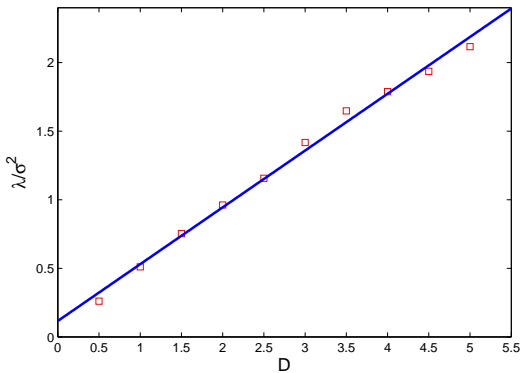}}
\caption{\label{geomean}The geometric mean of the eigenvalues of
the covariance matrix as a function of the signal strength $D$.
The solid line is a linear fit. The size of the system is $L=150$
and the step size is $0.06$. }
\end{figure}

The entropy (per symbol) of the noise in the Gaussian approximation is:
\begin{equation}
H[\delta z] = \ln \left((2\pi e)^{\frac{1}{2}}\bar{\lambda}\right)
\label{noise_entropy}
\end{equation}
where $\bar{\lambda}$ is the geometric mean of the eigenvalues of
the ($2N\times 2N$, $N$ being the number of eigenvalues) covariance matrix
of  $\delta\xi_i,\delta\eta_i$, $C_{ij}$. In FIG.~\ref{geomean} we see a plot of $\bar{\lambda}\over{\sigma^2}$
for ten values of $D$.

The linear fit gives $\bar{\lambda} = 0.41 \sigma^2 D$, and allows us
to compute a first result for the spectral efficiency lower bound (again, ignoring jitter):
\begin{equation}
R = \ln \left(\left(\frac{\pi}{8e}\right)^{\frac{1}{2}}\frac{BG\ln G}{0.41h\nu_0\eta_{sp}(G-1)^2}P_ct_c^2\right)
\label{lower_bound}
\end{equation}
where  $P_c,t_c$ are the characteristic units of power and time used to normalize
the units of the NLSE and they are inserted here to express the result
in SI units. For a system with $N_a$ amplifiers, one needs to replace
$\sigma^2$ with $N_a \sigma^2$. Plugging in typical values for the quantities involved
($P_c=0.05W,t_c=3\times10^{-11}s,G=100,\eta_{sp}=2,
h=6.6\times10^{-34}J\cdot s,\nu_0=2\times10^{14}Hz,N_a=10$) we find:
\begin{equation}
R = \ln\left(7\times10^{-7}\times B\right)\; \mbox{bits/s/Hz}
\end{equation}
with the bandwidth $B$ measured in Hz. Using all the available
bandwidth for optical fibers, $B = 50 \mbox{THz}$ this yields a
value of $R \simeq 25 \mbox{bits/s/Hz}$, which is significantly
higher than the predicted result for the case of WDM
\cite{Mitra_Stark_2001}. One thing to note about this result is
that $D$ altogether disappears, cancelling between the incoming
entropy and the noise. This is because the variance of the
eigenvalue shift is linear in $D$. This is in contrast to
analogous results in WDM models, where increasing the signal
strength also increases the multiplicative noise between channels
and the capacity falls after reaching a maximum for a finite value
of the SNR \cite{Mitra_Stark_2001,Stark_Mitra_Sengupta_2001,
Narimanov_Mitra_2002,Green_et_al_2002_XPM,Kahn_Ho_2004_2004_IT_review}.

However, this result is incomplete as long as it does not include
the effects of jitter, which could significantly reduce the
capacity. To do this we need to consider not just the eigenvalues
$\{z\}$, but also the $\{\ln b_z\}$, so that the signal
description is complete \footnote{Another approach would be to use
a coding scheme where the eigenvalue space is partitioned into
bins and codewords are formed from the number of solitons in each
bin. This approach does not enjoy the same generality since it
posits a specific coding scheme, but it is nevertheless
interesting and analytically tractable and may become the subject
of future work.}. The evolution of $\ln b_z$ is given by
\cite{Konotop_book}:
\begin{eqnarray}
\begin{array}{l}
\frac{\partial\ln |b_z|}{\partial x} = 8\xi\eta \\ \\
\frac{\partial\phi}{\partial x} = 4(\eta^2 - \xi^2)
\end{array}
\label{more_noise}
\end{eqnarray}
so that $\ln b_z$ is affected by the noise in $\xi$ and $\eta$. There
is also noise specific to $\ln b_z$ at each amplifier, but this
is bounded and can be neglected for long-range transmission. To properly include
the $\{\ln b_z\}$ in our scheme and get a reliable lower bound for the capacity
we must ascertain their correlations, which we have found
to be non-zero. We are currently attempting to find an analytic expression for the correlations.
If this proves impossible, we shall compute their correlation
matrix numerically starting from (\ref{lnb_evolution}) and use it to compute
the incoming entropy. Since the $\{\ln b_z\}$ follow a Gaussian distribution,
this does not spoil the lower bound feature.


\end{document}